\documentstyle[aaspp4]{article}

\begin{document}     
\title{ Transient Emission From Dissipative Fronts in Magnetized, Relativistic
Outflows. II. Synchrotron Flares}

\author{Amir Levinson}
\affil{School of Physics and Astronomy, Tel Aviv University, 
Tel Aviv 69978, Israel}

\begin{abstract}

The time dependent synchrotron emission from relativistic jets,
and the relation between the synchrotron and ERC emission is 
considered within the framework of the radiative front model.
The timescale and profile of the optically thin emission are 
shown to be determined, in this model, by the shock formation radius, 
the thickness of expelled fluid slab and the 
variation of the front's parameters due to its transverse
expansion.  For a range of reasonable conditions, a variety of 
flare shapes can be produced, varying from roughly symmetric
with exponential rises and decays, as often seen in blazars, 
to highly asymmetric with a fast rise and a much slower, power law 
decay, as seen in GRB afterglows.  
The onset, duration, and fluence of low-frequency (below the initial 
turnover frequency) and hard gamma-ray (above the initial gamma-spheric
energy) outbursts are limited by opacity effects; the emission at 
these energies is quite generally delayed and, in the case 
of sufficiently short length outbursts, severely attenuated. 
The observational consequences are discussed.
One distinctive prediction of this model is 
that in a single, powerful source, the upper cutoff of the gamma-ray 
spectrum should be 
correlated with the timescale of the outburst and with the amplitude
of variations at long wavelengths (typically radio to millimeter). 

\end{abstract}

\section{Introduction}

It is widely believed that the rapid variability frequently observed 
in blazars (for reviews see Ulrich, et al. 1997; Wagner, 1997) is 
linked to dissipation in relativistic jets.  The 
characteristically short variability time scales 
-  down to a few minutes at optical and x-ray 
wavelengths (e.g., Wagner 1997), and several hours 
at gamma-ray energies (Mattox, J., et al. 1997;  Aharonian et al. 
1998) - indicate that the source is compact and strongly beamed, 
supporting this view.
Despite the apparent complexity of the temporal behavior exhibited by
blazars, ongoing observational efforts have revealed some
trends.  Firstly, the high-and low-energy emission appears to be 
well correlated.  Correlations between optical/UV and gamma-ray 
emission have been observed
in many blazars (Wagner, 1997, and references therein), and 
between the X-ray and TeV emission in the TeV BL Lac objects 
(Macomb et al. 1995; Buckley et al. 1996; Aharonian et al. 1998).
There is also evidence for a correlated activity at radio and gamma-ray
bands in some cases (e.g., 3C279; Wehrle et al. 1998).  
Secondly, the variation of the synchrotron emission below the spectral
turnover has the tendency to be slower and later at longer wavelengths
(with the exception of radio IDV), suggesting that the low frequency 
(radio-to-IR) emission is self-absorbed at the early stages of 
the outburst.  In spite of these trends, it seems that blazars can 
exhibit a variety of 
temporal behaviors, and that even in individual sources the variability
pattern can change, as exemplified by observations of the BL Lac 
object PKS 2155-304 that revealed different variability patterns 
during the 1991 (e.g., Brinkmann et al. 1994; Edelson et al. 1995) 
and 1994 (Urry et al. 1997) 
campaigns.  The change in temporal behavior can be ascribed to the 
different physical conditions that exist in the source during each 
episode, and it is, therefore, 
desirable to examine the dependence of the temporal characteristics 
of the emission on the physical parameters of the source.

In this paper we study the time dependent synchrotron emission from
relativistic jets in the framework of the radiative front model developed 
previously to study gamma-ray flares (Levinson, 1998; hereafter Paper 
I).  The model is extended to encompass synchrotron emission
and employed to calculate the temporal evolution of the synchrotron 
and ERC emission radiated by the front, with an attempt to elucidate 
the dependence of the variability pattern on the model parameters.
As shown below, the model can successfully account for some of the trends 
mentioned above, and predicts some additional features.  In \S 2 we 
give an outline of the model and the necessary extensions.  In \S 3 we
derive the equation governing the evolution of the synchrotron flux
in the front, and explain how it is incorporated into the numerical 
model.  We then go on to describe the numerical results.  The 
observational consequences are discussed in \S 4.

\section{Description of the model}

The model presented in Paper I assumes that the variable 
emission seen in blazars originates inside dissipative fronts 
that are produced by overtaking collisions of highly 
magnetized, relativistic outflows, and consists of a pair of 
shocks and a contact discontinuity (Romanova \& Lovelace, 
1997; Levinson \& Van Putten, 1997).  The front propagates 
with a velocity intermediate between that of the slow and 
fast fluid slabs, dissipates a fraction of the outflow energy, 
and cools radiatively and adiabatically.  A fraction of the 
dissipation power is taped for the acceleration of electrons to 
high energies, and the rest for the heating of the bulk plasma in 
the front.  The acceleration time of non-thermal electrons 
has been assumed to be short compared with the cooling and 
the corresponding light crossing times.  Further, the fluid 
parameters and the fraction 
of dissipation power that is taped for electron acceleration 
are assumed to have no explicit time dependence (but do 
depend on time implicitly through the dynamics of the front).  As
a consequence, the characteristics of the transient emission 
produced by this model reflect essentially the dynamics of the 
system, as well as the intensity of external radiation, rather than    
explicit variations of the outflow parameters and/or electron 
acceleration rate, as, e.g., in the one-zone, homogeneous model 
by Mastichiadis \& Kirk (1997).

In Paper I we focused on the production of gamma-ray flares in 
the regime where ERC emission (e.g., Dermer \& Schlickheiser 1993;
Blandford \& Levinson 1995) dominates over 
SSC emission (Bloom \& Marscher 1993).  
The dynamics of the system and the time evolution of the flux 
radiated by the front have been calculated in a self-consistent 
manner, by numerically solving the MHD equations governing the front 
structure coupled to the kinetic equations describing the angle 
averaged pair and gamma-ray distribution functions.
The solutions depend on four key parameters: i) the 
dissipation rate of magnetic energy, ii) the maximum injection energy of
electrons, denoted by $E_{emax}$, which was taken to be fixed in the 
observer frame for simplicity, iii) the fraction of dissipation
energy that is injected in the form of non-thermal electrons, and iv)
the ratio of the thickness of ejected fluid slab and the gradient length
scale of background radiation intensity.  The latter parameter 
determines essentially whether the shape and timescale of the flare are 
related to the radial variation of ambient radiation intensity, 
in which case the light curve is asymmetric with a rapid rise
and a longer decay, or to the shock travel time across the fast 
fluid slab, in which case the decay time is of order the cooling time 
and is typically much shorter than in the case of infinite length 
outbursts.  As shown in Paper I, the temporal evolution of the 
gamma-ray flux is insensitive to the injected electron spectrum 
provided that electron acceleration is efficient, and
that $E_{emax}$ exceeds the corresponding gamma-spheric energy at
any given radius along the course of the front.  The evolution of
the synchrotron spectrum, on the other hand, is expected to depend 
on variations of $E_{emax}$, particularly if the front is optically 
thin at the corresponding frequencies (cf. Levinson 1996). 
We shall, therefore, relax the assumption that $E_{emax}$ is fixed.
In general, the evolution of the injected electron distribution 
is governed by the acceleration process, and a self-consistent 
treatment of particle acceleration is required in order to follow 
the development of $E_{emax}$.  Such a treatment is beyond the scope 
of our paper.  Here we settle for a simple prescription where
the injected spectrum is taken to be a power law with an exponential 
cutoff above $E_{emax}(t)$, where
\begin{equation}
E_{emax}(t)=E_o (r/r_o)^{b}, 
\label{eq:Emx}
\end{equation}
with $r(t)=r_o+\beta_c ct$.  

One important feature of the model is a positive radiative feedback
that affects the emitted spectrum considerably.  The point
is that as a result of radiative losses the front decelerates and 
its expansion rate decreases until the peak of the radiated
power is reached.  The deceleration of the front leads, in turn, 
to enhanced dissipation rate of the bulk flow energy, since 
the relative velocity between the fast fluid and the reverse 
shock increases.  After peak emission is reached, the front
begins to reaccelerate, ultimately reaching its initial 
speed and structure.  
The consequences of this radiative feedback for the variability 
have been discussed in detail in Paper I.  Another
consequence is that the variability should depend strongly
on the orientation of the source, owing to the change in beaming 
factor during the deceleration phase.  The analysis presented in Paper I 
cannot account for such orientation effects, as it only treats the 
angle averaged flux.  The model does describe rather well, 
however, the variability that would be seen by an observer at 
sufficiently small viewing angle.  A study of orientation 
effects will be presented elsewhere (Eldar \& Levinson, in preparation).
                                                                
\section{Synchrotron flares}

In this section
we extend the analysis to incorporate synchrotron emission, and
apply the results to study the relation between the low-and 
high-energy emission in ERC dominated blazars.  In order to do
so, we augment the set of equations introduced in Paper I by an
equation governing the time change of the synchrotron intensity
(see eq. [\ref{eq:transfer}] below).  We also include additional loss 
term for the electrons accounting for synchrotron cooling.  The equations
are then integrated in the injection frame (the rest frame of the
boundary from which the fluid is expelled), as described in Paper I.
Now, the synchrotron opacity depends on the electron density and magnetic
field inside the front and, consequently, on the transverse expansion 
of the front.
Let $c_{s\perp}$ denotes the expansion speed in the transverse direction,
$d$ the cross-sectional radius, and $A=\pi d^2$ the cross-sectional area.
The expansion speed $c_{s\perp}$ may depend on external 
pressure and magnetic fields, or the density of ambient gas if inertial
effects are important, and should not be the same as the sound speed
in the radial direction in general.  We suppose that initially $d_o=
\psi r_o$, where $\psi$ is the jet opening angle, and $r_o$ is the 
radius of formation of the front (Paper I).  Then $d=d_o+c_{s\perp}t=
\psi[r_o+(c_{s\perp}/\psi)t]$.  For illustration we take $
(c_{s\perp}/\psi)$ to be equal to the front velocity, $c\beta_c$.  
We then obtain  $A(t)\propto r^2(t)$, where $r(t)=(r_o+c\beta_c t)$.
Let $N_e$ be the total number of electrons in the front.  The 
corresponding number density is then given by $n_e=N_e/(\Delta X A)$,
where $\Delta X$ is the axial length of the front.
The magnetic field is assumed to decline like 
\begin{equation}
B=B_o [r(t)/r_o]^{-p}. 
\label{eq:B} 
\end{equation}

\subsection{Transfer equation}

We define $I_{\nu}(t,\mu)$ to be the unpolarized synchrotron intensity 
inside the front, as measured in the injection frame.  
The equation governing the time evolution of $I_{\nu}$ can be derived
in the manner described in Paper I.  One then obtains,
\begin{equation}
\frac{\partial}{\partial x^0}I_{\nu}=j-[(\mu-\beta_{s-})/
(\Delta X)+\kappa] I_{\nu},
\label{eq:transfer}
\end{equation}
where $j(\nu,\mu,t)$ and $\kappa(\nu,\mu,t)$ are the emission 
and absorption coefficients, $x^0=ct$, $\beta_{s-}$ is the velocity 
of the reverse shock, $\cos^{-1}\mu$ is the angle between the direction
of a photon and the front velocity, and $\Delta X$ is the axial 
length of the front.  The term proportional to $(\mu-\beta_{s-})$
accounts for the change in the intensity due to the combined effects
of front expansion and escape of synchrotron photons from 
the front (see Paper I for details).  For convenience, we compute 
the emission and absorption coefficients in the rest frame of 
the front (quantities 
measured in the front frame are henceforth denoted by prime), and then
transform them to the injection frame.  The coefficients in the two 
frames are related through (Rybicki \& Lightman 1979), 
\begin{eqnarray}
j(\nu,\mu)=\left(\frac{\nu}{\nu'}\right)^2j'(\nu',\mu'), \\
\kappa(\nu,\mu)=\frac{\nu'}{\nu}\kappa'(\nu',\mu'),
\end{eqnarray}
with $\nu={\cal D}\nu'$, ${\cal D}=[\Gamma_c(1-\beta_c\mu)]^{-1}$ being
the Doppler factor, where $\Gamma_c$ is the bulk Lorentz 
factor of the front.  The emission and absorption coefficients are 
given in the front frame by,
\begin{equation}
j'=\frac{\sqrt{3}e^3}{4\pi mc^2}B(t)\sin\phi'\int{n'_e({\cal E}',t)
F(x') d{\cal E}'},
\end{equation}                                                          
\begin{equation}
\kappa'=\frac{\sqrt{3}e^3}{8\pi m \nu'^2}B(t)\sin\phi'\int{{\cal E}'^2
\frac{d}{d{\cal E}'}\left[\frac{n'_e}{{\cal E}'^2}\right]
F(x') d{\cal E}'}.
\label{eq:kp}
\end{equation}
Here ${\cal E}'$ is the electron energy with respect to the front frame, 
$n'_e$ is the corresponding electron number density per unit energy, 
$\phi'$ is the angle between the line of sight and the magnetic field
in the front frame,
$x'=\nu'/\nu'_c$; $\nu'_c=(3e/4\pi mc)B(t)\sin\phi' ({\cal E}'/mc^2)^2$,
and $F(x)=x\int_{x}^{\infty}K_{5/3}(\chi)d\chi$. 
We emphasize that $B(t)$, $\Gamma_c(t)$, and $\beta_{s-}(t)$ are 
not given explicitly but rather calculated self-consistently by 
integrating the front equations.    
Now, as mentioned above, the integration of the electron kinetic 
equation is carried out in the injection frame.  This yields $n_e(
t,{\cal E})$.  Thus, in order to obtain $n'_e(t,{\cal E}')$ we need to 
transform $n_e$ into the front frame.  The appropriate 
transformation is derived in the Appendix under the assumption
that the comoving electron distribution is isotropic.  

As a check on the above equation, consider the case of a radiating blob,
for which $\beta_{s+}=\beta_{s-}=\beta_{c}$.  It can be readily 
shown that the steady-state solution 
of eq. (\ref{eq:transfer}) reduces, in the optically thin limit,  
to $I=j\Delta X/(\mu-\beta_c).$  On
substituting $\Delta X'=\Gamma_c\Delta X$ and $\mu'=(\mu-\beta_c)/
(1-\beta_c\mu)$ in the latter expression, we recover the well known 
result: $I={\cal D}^3(j'\Delta X'/\mu')
={\cal D}^3 I'$.  Note that $\Delta X'$ and $\mu'$
are respectively the blob thickness and the cosine of the 
angle between the blob velocity and the direction of a photon, 
as measured in the blob frame.

We shall proceed by assuming that the 
magnetic field in the front is tangled and has no net 
direction.  The emission and absorption 
coefficients can then be averaged over $\phi'$.  Since 
we do not consider polarization and orientation effects,
we anticipate the results not to depend strongly on this 
assumption.  We further average eq. (\ref{eq:transfer}) over 
the viewing angle $\mu$.

\subsection{Cooling rates}

The synchrotron cooling time can be expressed in terms of the 
front parameters as
\begin{equation}
\frac{ct_{syn}}{r_o}=0.17\left(\frac{L_{j46}}{r_{o16}}\right)^{-1}
\Gamma_{A-}^2(n_f/n_{-})U^2_{Af}({\cal E}/mc^2)^{-1}.
\end{equation}
Here $n_f$, $n_{-}$ are the electron densities inside the
front and in the fluid exterior to the front, respectively, 
$U_{Af}$ is the Alfv\'en 4-velocity with respect to front frame,
$L_{j46}$ is the power carried by the fast outflow in units of 
$10^{46}$ erg s$^{-1}$, $\Gamma_{A-}$ is the Lorentz factor associated
with the Alfv\'en 4-velocity of the exterior fluid, and $r_{o16}$ is 
the radius of shock formation $r_o$ in units of $10^{16}$ cm.  
For magnetized jets $\Gamma_{A-}>>1$.  The product ($n_f/n_{-})U_{Af}$ 
changes with time due to the radiative feedback (Paper I), and is 
typically of order 30 or so in the presence of rapid magnetic field 
dissipation.  Thus, the synchrotron cooling time 
at most electron energies is much shorter than the dynamical time
$r_o/c$.  
The ratio of synchrotron and Compton cooling rates equals the ratio 
of comoving energy densities of magnetic field and external radiation
field, and is given by
\begin{equation}
\frac{t_{IC}}{t_{syn}}\simeq \frac{20}{\Gamma_{A-}^2\Gamma_c^2}
\frac{L_{j46}}{(\epsilon L_{s})_{45}}(n_f/n_{-})U^2_{Af}
\label{eq:cool}  
\end{equation}
with $(\epsilon L_{s})_{45}$ being
the fraction of soft-photon luminosity that is scattered across the 
jet in units of $10^{45}$ erg s$^{-1}$.  This ratio is independent of
radius when $p=1$ in eq. (\ref{eq:B}) (provided the soft photon 
intensity declines as $r^{_2}$).  Nevertheless, enhanced compression 
during the radiative phase renders it time dependent, as shown in fig. 1. 
For the parameters adopted in fig. 1 electron cooling is dominated 
by ERC emission as long as $L_j<10(\epsilon L_{s})$.

\subsection{Numerical results}

The numerical model presented in Paper I has been extended to include
equations (\ref{eq:transfer})-(\ref{eq:kp}).  The entire set of 
equations has then been integrated, starting from $r_o$ where 
$I_{\nu}=0$, and where the front structure is taken to be that of 
an adiabatic front (see Paper I for a detailed discussion concerning 
the initial conditions).  In the following examples the Lorentz 
factors of the fluids ahead and behind the front and the rest frame
Alfven 4-velocity have been chosen to be respectively 5, 20, and 10,
as in Paper I.  A rapid magnetic field dissipation with the 
same decay constant as in Paper I ($\alpha_b=0.5$), and a background
radiation field with the same intensity have been invoked.

We examine first the dependence of the variability pattern on the 
thickness of expelled fluid slab, $d$, in the case in which the maximum 
injection energy, $E_{emax}$, is time independent.  As shown below,
both the shape of the light curve and the spectral evolution
depend on $d$.  Quite generally we find that 
the ejection of slabs of thickness $d>>r_o$ leads to the production
of asymmetric flares with a fast, exponential rise and a much 
slower, power law decline, as shown in the {\it bottom right} panel
of fig. 2, where a sample light curves of the optically thin emission, 
computed for different values of $d/r_o$ is exhibited .  The decay 
time and profile in this case are determined predominantly by 
the decline of the density and magnetic field inside the front due 
to its transverse expansion.  When the expelled slab is sufficiently 
thin, specifically, when $d/r_o$ is smaller than approximately 
$(1-\beta_c)$, $\beta_c$ being the average front velocity, shock 
crossing of the fluid slab followed by a large drop of the energy 
dissipation and consequent particle acceleration rates occurs on
a timescale much shorter than the time change of the front parameters
due to expansion.  As a result, the radiated flux decay on the cooling
time scale (provided it is short enough), leading to a roughly 
symmetric light curve with exponential rise and decay.  The observed 
duration of the flare in this case is on the order of $d/c$.  An example is 
shown in the {\it upper left} panel in fig. 2.  Intermediate length 
outbursts can give rise to light curves that exhibit a fast rise, a 
plateau and a steep decline, as illustrated in the {\it upper right} 
and {\it bottom left} panels.

Fig. 3 displays light curves at different observing frequencies 
(10, $10^3$, $10^5$, and $10^6$ GHz, with logarithmic energy intervals),
produced for $d/r_o=10^{-2}$, and different values of $r_o$.
As seen, the low frequency flux is delayed, owing to self absorption
at the early stages of the outburst.  Furthermore, the low frequency
emission is strongly depressed when the radius of front formation, $r_o$,
is sufficiently small.  In the example depicted in fig. 3 there
is essentially no radio emission for $r_o$ smaller than about 10, and
very little emission even at millimeter wavelengths for $r_o<1$. 
The suppression of the flux at $10^6$ GHz seen for $r_o=10^2$ in the
{\it upper left} panel in fig. 3, is due to our choice of $E_o$ and
$B_o$ in eqs. (\ref{eq:Emx}) and (\ref{eq:B}).  The total gamma-ray flux
is emitted from roughly the same radii as the optically thin 
synchrotron flux and, therefore, correlations between the gamma-ray and 
optical/UV emission are expected.  The precise relation, however, depends
upon the combination of parameters through the radiative feedback and 
the evolution of the pair cascade, but the time lags should not exceed 
the flare duration.  We emphasize that the gamma-ray flare can either lead
or precede the optical/UV emission.  As discussed in Paper I, the gamma-ray 
flux above the initial gamma-spheric radius should also be 
delayed and, for small enough $r_o$, will be strongly attenuated by 
pair production on external photons.  Hence, we anticipate 
some correlated activity of the low-frequency and the hard gamma-ray 
emission (see further discussion below).

The dependence of flare's properties on $p$ (defined in eq. 
[\ref{eq:B}]) is shown in fig. 4, for infinite length outbursts.  As 
expected, increasing $p$ results in a substantially steeper decay 
of the emitted flux.  As illustrated by this example, roughly 
symmetric flares can be produced essentially 
for a steep enough magnetic field profile, even in the case of thick 
slabs.   The suppression of the low frequency flux at larger values of $p$, 
seen in the figure, is due to the steeper decline of the ratio 
of synchrotron and Compton cooling rates with time (see eq. [\ref
{eq:cool}]).  In the case of short length outbursts, the light curve 
is essentially insensitive to the radial profile of the magnetic field. 

Finally, we consider a situation whereby the maximum injection energy 
evolves with time.   We then anticipate the development of the 
synchrotron flux at frequencies above the initial cutoff frequency,
$\nu_{max}=(3eB/4\pi m_ec)E_o$ (cf. eq. [\ref{eq:Emx}]), to be governed 
by the evolution of $E_{emax}(t)$,
whereas at lower frequencies the light curves should be highly 
insensitive to  $E_{emax}(t)$, reflecting mainly 
the evolution of the synchrotron opacity, as in the previous example.
This is illustrated in fig. 5, where light curves computed for 
infinite length outbursts and $b=1$ in eq. (\ref{eq:Emx}) are 
plotted.  As seen, the onset of the flare at high frequencies becomes 
more delayed and the maximum flux becomes smaller for lower cutoff 
energies, $E_o$. 

\section{Summary and Conclusions}

In this paper we considered the time dependent synchrotron emission
produced by radiative fronts propagating in a magnetically dominated 
jet, and examined the relation between 
the synchrotron and ERC emission.  Using a numerical model developed 
earlier, we analyzed the dependence of the variability pattern on the 
model parameters, and demonstrated how a change in physical parameters
can lead to considerably different temporal behavior.
We now summarize our results and conclusions.

Two important parameters determine, in addition to the radial profile
of the jet's magnetic field, the shape of the flare and the 
spectral evolution of the broad-band emission; the thickness of 
expelled fluid slab, and the radius at which the front is created.
Ejection of sufficiently thin slabs leads to the production
of roughly symmetric flares with exponential rise and decay, as 
occasionally seen in blazars (e.g., Massaro et al. 1996; 
Urry et al. 1997).  The duration of 
the optically thin flare in this case should be on the order of the 
light crossing time of the ejected fluid slab, which can be as short 
as $r_{g}/c$, where $r_{g}$ is the gravitational radius of the putative 
black hole (cf. Paper I).  The low frequency emission is quite generally 
delayed due to optical depth effects.  Such delays provide an important 
diagnostic of radiatively efficient, inhomogeneous models in general, and
are in agreement with the delays (of weeks to months) between high-energy 
and radio outbursts often observed (Reich et al. 1993; Wehrle et a.l, 1998), 
and the ejection of a superluminal blob following a gamma-ray flare 
(Zhang et al. 1994; Wehrle et al. 1996; Otterbein et al. 1998).  The  
much shorter lags (hours to days) between the peaks of the gamma-ray, UV 
and optical fluxes, as seen on several occasions (e.g., Edelson et al. 1995; 
Urry et al. 1997; Wehrle et al., 1998), 
are also a natural consequence of this model.  The precise relation 
between the optically thin emission in different bands depends on 
the conditions in the source through the radiative feedback and the 
evolution of pair cascades inside the front.
If, in addition to being thin, the front is formed at a sufficiently 
small radius, then 
the low-frequency synchrotron flux will be strongly suppressed by virtue 
of self absorption, and the gamma-ray flux at energies above the 
corresponding gamma-spheric energy, will be severely attenuated
by pair production on external photons ahead of the front.
This implies that i) during such episodes a 
source may exhibit gamma-ray and UV/optical outbursts followed by
little or no activity at low (typically radio-to-submillimeter) 
frequencies, and ii) in a single object the cutoff energy of 
the gamma-ray spectrum should be smaller for shorter outbursts, 
and should be correlated with the lack of activity (or with 
the amplitude of variations) at long wavelengths.  Successful 
detection of such 
correlations in a source, particularly ii), will provide a strong 
support to this model, as such a behavior is not expected in 
SSC models or other types of inhomogeneous models and is, therefore, 
distinctive.  For a reasonable choice of parameters
we estimate the cutoff energy of the gamma-ray spectrum in the 
powerful blazars (which are likely to be ERC dominated) to lie
in the range between a few GeV to a few hundreds GeV.  It may be 
possible to observe it with the next generation gamma-ray
telescope or, perhaps, with upcoming ground based experiments. 

Intermediate length 
outbursts can produce synchrotron light curves that exhibit a relatively 
rapid rise, a plateau and a sharp decay.  We argue that the 
light curves reported by the PKS 2155, 1991 campaign 
(Urry et al. 1993; Courvoisier et al. 1995) 
might have been produced by such an event.  A rise on a time scale of 
order a few days followed by a period of about 20 days during which 
the average flux remained at maximum level is evident (although the flux 
fluctuated around the maximum level during this period).  Unfortunately,
there is no data at later times.  The faster, small amplitude oscilations
may be caused by subsequent formation of smaller fronts, by instabilities,
or by inhomogeneities in the upstream flow (which would lead essentially 
to bifurcation of the front).  Furthermore, there is indication 
(Edelson et al. 1995) that 
the variation of the radio emission is delayed and has a smaller 
amplitude, consistent with this model.  It is also tempting to relate
the difference in temporal behavior seen by the 1991 and 1994 campaigns
to the difference in conditions, particularly the length of the outburst, 
during those two episodes (cf. Georganopoulos \& Marscher 1997).  

When the thickness of expelled fluid slab largely exceeds the formation 
radius, which happens when the injection time of the jet by the central 
engine is much longer than the fluid acceleration time, the resultant 
light curve will exhibit an exponential rise and a
power law decline.  If the radial decrement of the jet's magnetic field 
is not too steep ($p<2$ in ep. [\ref{eq:B}]) then the flare will 
appear asymmetric, with a decay much longer than the rise when viewing
the source at small enough angles.  At larger viewing angles the shape 
of the flare may be altered by orientation effects (Eldar \& Levinson,
in preparation).  Such asymmetric flares are atypical to blazars,
but are characteristic to GRB afterglows. 

In situations where the maximum injection energy increases with 
time, the onset of outbursts at frequencies above the initial upper 
cutoff becomes delayed, with longer time lags and smaller 
peak fluxes at higher frequencies, in contrast to the tendency
found in the case of time independent $E_{emax}$.  The gamma-ray light
curve should not be affected significantly though (Levinson 1996), in
contrast to the evolution predicted by SSC type models.  Such 
events can lead to 
delayed, lower amplitude variations at optical/UV wavelengths, or even
to the lack of apparent variations at these frequencies.  Since the 
evolution of $E_{emax}$ is dictated by the rate of electron acceleration
in the front, careful examination of the correlations at short 
wavelengths can provide important information regarding the 
acceleration process. 

Finally, we note that the inclusion of the SSC process 
may alter our results somewhat, particularly for parameters
typical to fainter sources.  The study of SSC flares is left
for future work.

I thank Avigdor Eldar for useful discussions, and the anonymous 
referee for constructive criticism.
This research was supported by a grant from the Israeli Science Foundation
and by Alon Fellowship.
    
\appendix
\section{Appendix: Transformation of the electron distribution function}
                                                                        
We suppose that the electron distribution is isotropic in the frame of
the front, and denote by $n({\cal E},\mu)$ and $n'[{\cal E}'({\cal E},\mu)]$
the differential number density per unit energy per solid angle, as measured
in the injection frame and front frame, respectively, where
\begin{equation}
{\cal E}'({\cal E},\mu)={\cal E}\Gamma_c(1-\beta_e\beta_c\mu);\ \ \ \ \beta_e=
\sqrt{1-{\cal E}^{-2}}.
\end{equation}
The relation between the electron distributions in the two frames reads
\begin{equation}
n({\cal E},\mu)=\left(\frac{{\cal E}}{{\cal E}'}\right)^2 n'({\cal E}').
\end{equation}
Averaging over angles yields
\begin{equation}
n({\cal E})=\int_{-1}^{1}n({\cal E},\mu)d\mu=\frac{{\cal E}}
{\Gamma_c\beta_c\beta_e}\int_{{\cal E}\Gamma_c(1-\beta_e\beta_c)}^
{{\cal E}\Gamma_c(1+\beta_e\beta_c)}\frac{n'({\cal E}')}{{\cal E}'^2}
d{\cal E}'.
\end{equation}
Taking the derivative with respect to ${\cal E}$ of the last equation
yields, 
\begin{equation}
n'[{\cal E}\Gamma_c(1-\beta_e\beta_c)]=
n'[{\cal E}\Gamma_c(1+\beta_e\beta_c)]\frac{(1-\beta_e\beta_c)^3
(\beta_e+\beta_c)}{(1+\beta_e\beta_c)^3(\beta_e-\beta_c)}
+\frac{\Gamma_c^3\beta_c(1-\beta_e\beta_c)^3}{(\beta_e-\beta_c)}
\left\{(3\beta_e^2-1)n({\cal E})-\beta_e^2\frac{dn({\cal E})}
{d\ln{\cal E}}\right\}.
\end{equation}
The distribution $n'({\cal E}')$ can be calculated now by recursion using
the last equation.
   
\break

\break

\begin{figure}
\vspace{10cm}  
\includegraphics{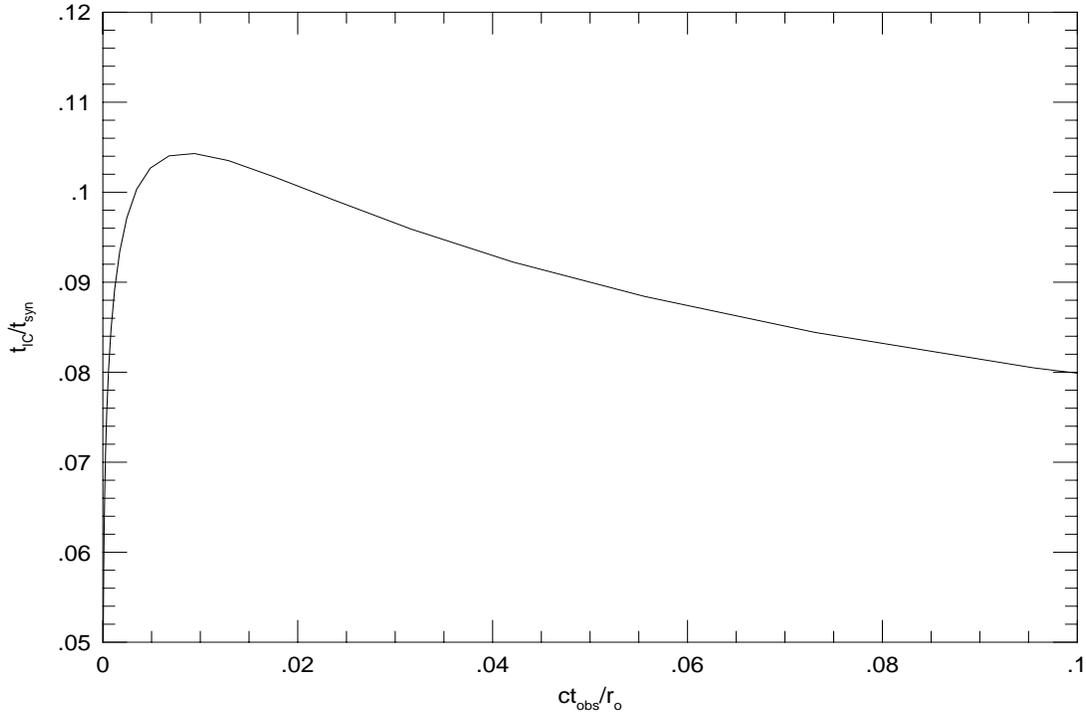}
\caption{Ratio of inverse Compton to synchrotron cooling times
as a function of observer time, for $r_{o16}=1$, $(\epsilon L_{s})_{45}=1$,
and $L_{j46}=1$.}
\end{figure}

\begin{figure}
\vspace{10cm}  
\includegraphics{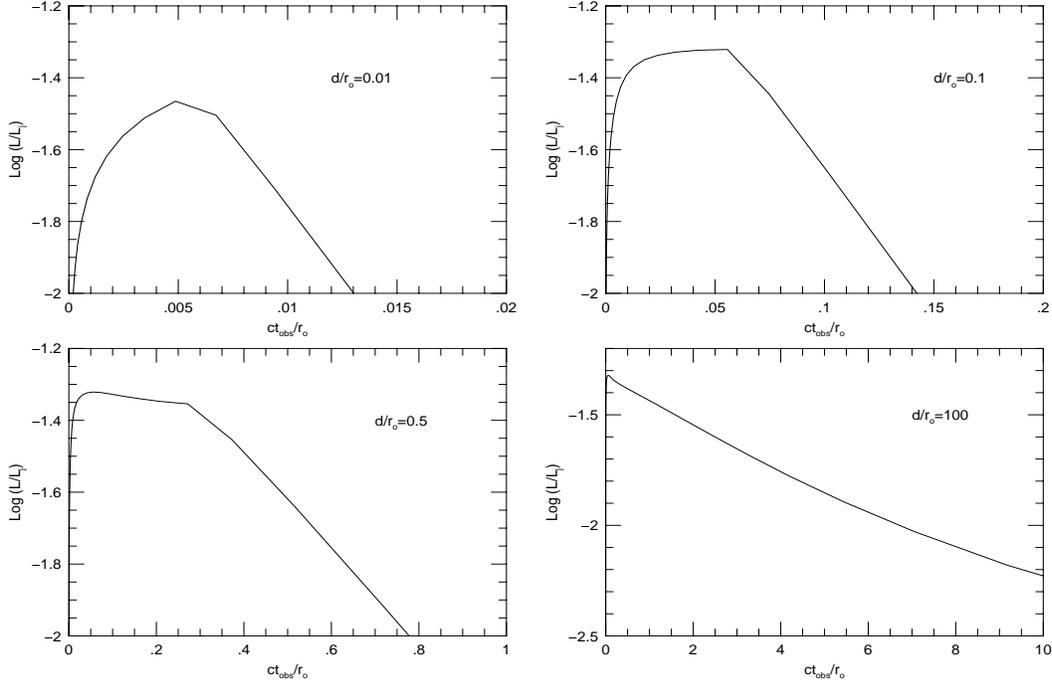}
\caption{Synchrotron light curves at $10^6$ GHz produced by
the model, for $L_{j46}=1$, $(\epsilon L_{s})_{45}=1$, $r_{o16}=1$,
and different values of $d/r_o$ (as indicated). 
Shown is the log of the apparent luminosity (in units of $L_j$) 
radiated into a logarithmic frequency interval centered at the 
corresponding frequency, as a function of observing time.}
\end{figure}

\begin{figure}
\vspace{10cm}  
\includegraphics{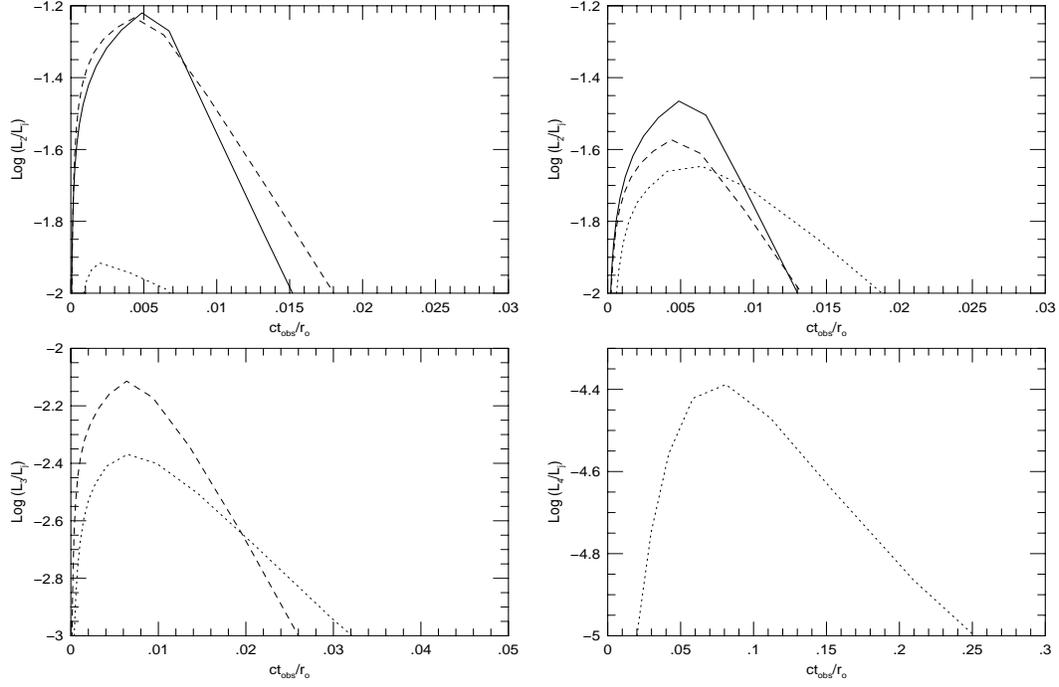}
\caption{Synchrotron light curves at 
$10^6$ GHz ({\em upper left}), $10^5$  GHz ({\em upper right}), 
$10^3$ GHz ({\em bottom left}), and $10$ GHz ({\em bottom right}), 
produced using $L_{j46}=1$, $(\epsilon L_{s})_{45}=1$, 
$d/r_o=10^{-2}$, and different values of $r_o$;
$r_{o16}=1$ (solid line), $r_{o16}=10$ (dashed line), and
$r_{o16}=10^2$ (dotted line).}
\end{figure}

\begin{figure}
\vspace{10cm}  
\includegraphics{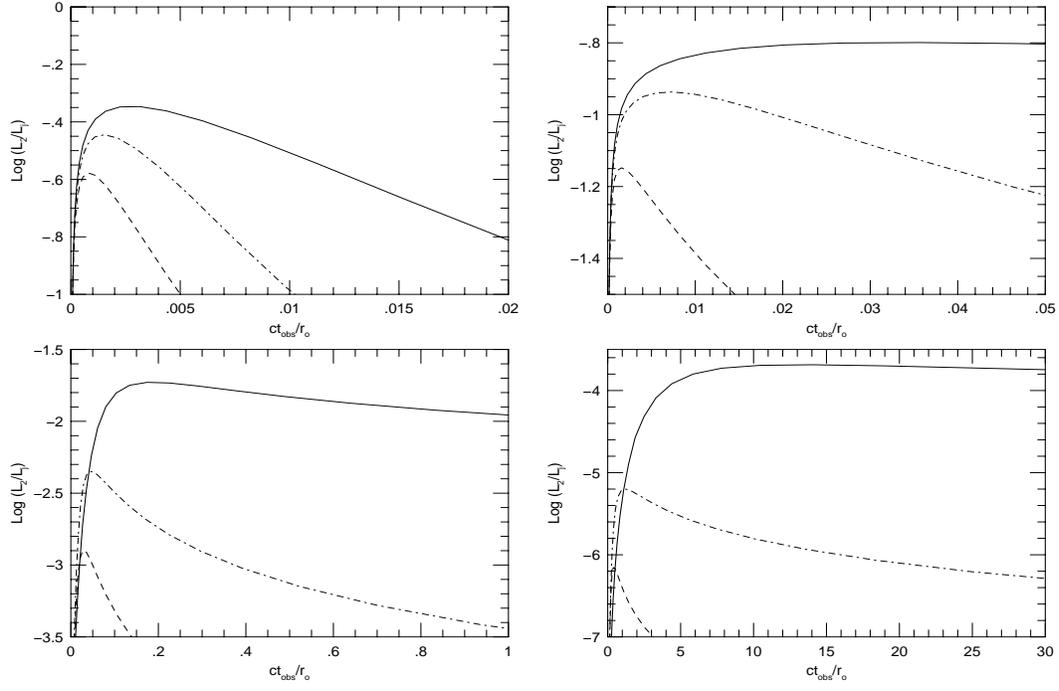}
\caption{Same as fig. 3, but for  $L_{j46}=1$, 
$(\epsilon L_{s})_{45}=0.1$, $r_{o16}=1$, $p=1$ (solid line), 
$p=1.5$ (dotted dashed line) 
and $p=2$ (dashed line).}
\end{figure}

\begin{figure}
\vspace{10cm}  
\includegraphics{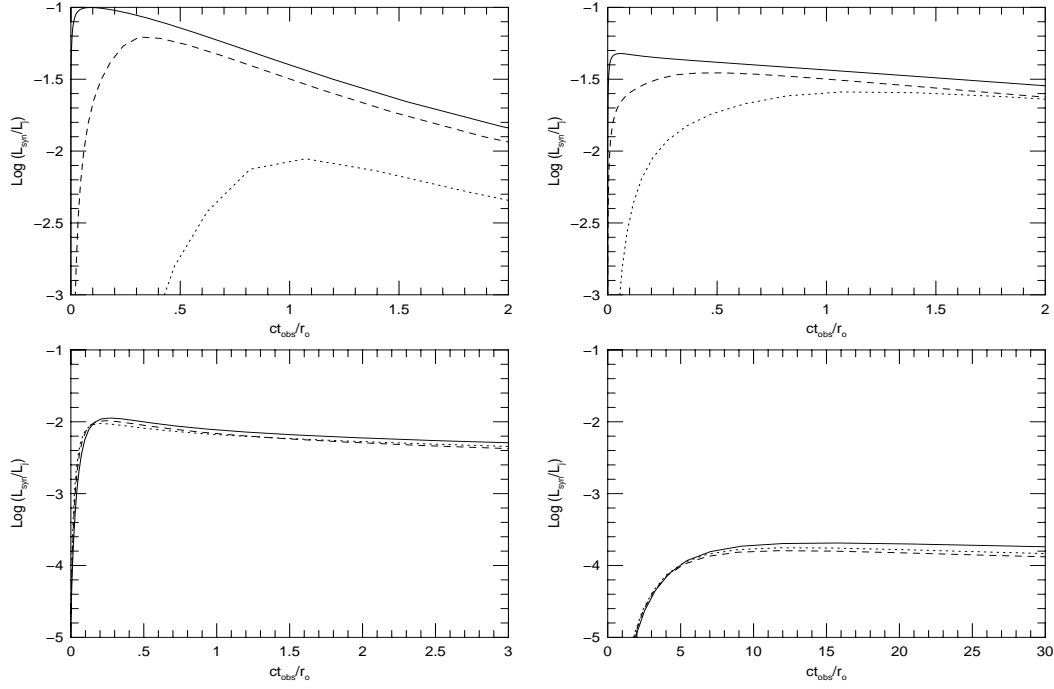}
\caption{Same as fig. 4 with $p=1$, but for injected electron spectrum 
with a maximum cutoff energy that varies according to eq. (\ref{eq:Emx}), 
with $b=1$, $E_o/(m_ec^2)=5\times10^2$ 
(dotted line), $2\times10^3$ (dashed line), and $10^5$ (solid line).}
\end{figure}
                                                                
\end{document}